\documentclass[%
  prb,%
  letterpaper,%
  twocolumn,%
  showpacs,%
  floatfix,%
  superscriptaddress%
]{revtex4-1}

\RequirePackage{graphicx,amsmath,amssymb,bm}
\RequirePackage{hyperref,verbatim}
\hypersetup{%
  breaklinks = {true},
  citecolor = {blue},
  colorlinks = {true},
  linkcolor = {red},
  pdfcreator = {\LaTeX\ and \flqq hyperref\frqq},
  pdffitwindow = {true},
  pdfmenubar = {true},
  pdfpagelayout = {SinglePage},
  pdfstartview = {Fit},
  pdftoolbar = {true},
  plainpages = {false},
}

\usepackage{siunitx}

\newcommand{\Fref}[1]{Fig.~\ref{#1}}

\newcommand{\Eqref}[1]{Eq.~(\ref{#1})}
\renewcommand{\eqref}[1]{Eq.~(\ref{#1})}

\usepackage{colortbl}
\makeatletter

    \def\CT@@do@color{%
      \global\let\CT@do@color\relax
            \@tempdima\wd\z@
            \advance\@tempdima\@tempdimb
            \advance\@tempdima\@tempdimc
    \advance\@tempdimb\tabcolsep
    \advance\@tempdimc\tabcolsep
    \advance\@tempdima2\tabcolsep
            \kern-\@tempdimb
            \leaders\vrule
                    \hskip\@tempdima\@plus  1fill
            \kern-\@tempdimc
            \hskip-\wd\z@ \@plus -1fill }
    \makeatother

\RequirePackage[usenames,dvipsnames]{xcolor}
\usepackage{soul}

\newcommand{\unicamplimeira}{Faculdade de Ci\^{e}ncias Aplicadas, Universidade
Estadual de Campinas, 13484-350 Limeira, SP Brazil}

\newcommand{\UCF}{ Department of Physics, University of Central Florida,
Orlando, FL 32816-2385, USA}

\begin{document}

\title{Disorder effect on the anisotropic resistivity of phosphorene determined by a tight-binding model}

\author{Carlos J. P\'aez}
\affiliation{\unicamplimeira}%
\author{Kursti DeLello}%
\affiliation{\UCF}%
\author{Duy Le }%
\affiliation{\UCF}%
\author{Ana L. C. Pereira}%
\affiliation{\unicamplimeira}%
\author{Eduardo R. Mucciolo}%
\affiliation{\UCF}%

\begin{abstract}
In this work we develop a compact multi-orbital tight-binding model
for phosphorene that accurately describes states near the main band
gap. The model parameters are adjusted using as reference the band
structure obtained by a density-functional theory calculation with the
hybrid HSE06 functional. We use the optimized tight-binding model to
study the effects of disorder on the anisotropic transport properties
of phosphorene. In particular, we evaluate how the longitudinal
resistivity depends on the lattice orientation for two typical
disorder models: dilute scatterers with high potential fluctuation
amplitudes, mimicking screened charges in the substrate, and dense
scatterers with lower amplitudes, simulating weakly bounded
adsorbates. We show that the intrinsic anisotropy associated to the
band structure of this material, although sensitive to the type and
intensity of the disorder, is robust.
\end{abstract}


\maketitle

\section{Introduction}

Two-dimensional (2D) materials formed by a few atomic layers are
likely to be featured in high-performing electronic devices in the
near future, thanks to their high charge mobility, strong gating
capabilities, and other unusual properties. For nearly a decade, the
focus was primarily on graphene,\cite{CasGP09} but its use in
transistors as a substitute for silicon has many limitations; in
particular, the absence of a bandgap.\cite{WenZ16}
The focus now has shifted to others 2D material. Among these,
monolayer black phosphorus, known as phosphorene, is particularly
attractive. Phosphorene has high charge mobility (typically $100-1000$
cm$^2$V$^{-1}$s$^{-1}$),\cite{LiYY14,LiuNZ14} its band gap spans a
wide range in the visible spectrum, and presents a strong in-plane
anisotropy.\cite{MisCY2015,XiaWJ14,LiuLR16}

Current methods for calculating the band structure and optical and
electronic properties of phosphorene include density functional theory
(DFT). Attempts at studying the electronic properties of phosphorene
have also been made using a self-consistent pseudopotential
approach.\cite{YukHA1981,RodCC2014,LowRC14,PaiHK06} Those approaches
are highly successful in predicting the overall trend of the band
structure, but they can be computationally expensive for calculating
the transport properties.

Previous works dealing with phosphorene focused on obtaining the
optical and electronic properties using tight-binding models with only
one $p_z$ orbital per
atom.\cite{RudYK2015,YuaRK2015,TagEZ2015,PopKS15} However, these
simple models do not capture the anisotropy in the electronic and
optical properties accurately.

Differently from graphene, the atomic layers in phosphorene are not
perfectly flat; instead, phosphorene has a puckered surface due to the
$sp^3$ hybridization. Thus, for an accurate description of the
electronic properties including the anisotropy, both $p$ and $s$
orbitals have to be taken into account.
Recently, a tight-binding model has been developed which includes
nearest and next-nearest neighbor interactions.\cite{Osada15} While
this model offers only a qualitative view of the behavior of the band
structure, it also provides reasonable predictions in agreement with
experimental results, and can serve as a good starting point for our
model.

In this work, we develop an effective tight-binding model for
phosphorene through a optimization procedure of the tight-binding
parameters. The tight-binding model is built with an orthogonal basis
composed of all 3$s$ and 3$p$ orbitals of phosphorus. It reproduces
very accurately the energy bands and reasonably well the orbital
compositions near the extremes of the conduction and valence bands, as
obtained by DFT calculations based on the hybrid HSE06
functional,\cite{Heyd2003, HSE2006} referred herein as DFT-HSE06.

Using this optimized tight-binding model, we calculate the linear
conductance of phosphorene for two different lattice orientation
(zigzag and armchair) in the presence of background potential
fluctuations that mimic disorder. Our aim is to investigate the
in-plane anisotropy in the transport when in presence of disorder. We
consider two limits of the Gaussian-correlated potential fluctuations:
low amplitudes with high density, and high amplitude with low
density. In both regimes, we find that the intrinsic anisotropy due to
the electronic structure is manifest in the resistivity of
phosphorene.

Phosphorene samples are shown to be very sensitive to the
environment,\cite{JosSD14,JosGH15,DogFK15} therefore, the role of
disorder represents an important issue, with both theoretical and
practical relevances.  First-principle studies of the effects of
vacancies,\cite{WeiY15} substitutional atoms,\cite{YuaFZ14}
oxidation,\cite{ZilCC15} and impurities\cite{KulMO15} have been only
carried out so far for small systems due to the high computational
cost. However, the computation of transport properties in particular
requires the carriers to be in the proper dynamical regime (diffusive
in most cases), which in turn can only be simulated in large enough
samples. Therefore, the influence of disorder on the transport
properties of phosphorene is not yet settled.

The remaining of the paper is organized as follows.
In Sec. \ref{sec:model}, the optimization procedure used in our
tight-binding model to calculate the band structure is presented.
In Sec.\ref{sec:bandstructure}, we compare the band structure obtained
from the DFT-HSE06 and from our optimized tight-binding model.
In Sec. \ref{sec:anisotropy}, the band structure around the
high-symmetry $\Gamma$-point is analyzed, allowing us to obtain
accurate values for the effective masses in zigzag and armchair
directions.  In Sec.\ref{sec:transport}, we study the effects of
disorder on the transport properties of phosphorene, specially on the
anisotropic resistivity. Finally, in Sec. \ref{conclusion}, we draw
our conclusions. The main text is supplemented by Appendix \ref{SK}
containing technical aspects of the simplified LCAO method
calculations.

\section{Model}
\label{sec:model}

\begin{figure} [hb!]
\includegraphics[width=1\columnwidth]{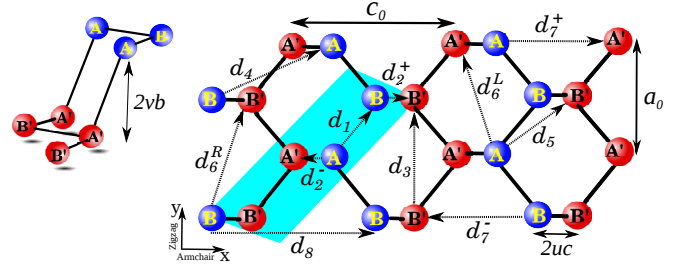}
\caption{(Color online)(a) Red (blue) circles represent phosphorus
  atoms in the lower (upper) layer. (b) Projection of the phosphorene
  crystal structure onto a two-dimensional plane. The rectangular area
  indicates the unit cell, which contains four phosphorus
  atoms. Zigzag and armchair edges are indicated.}
\label{fig:Fig1}
\end{figure}

The crystal structure of monolayer phosphorene is illustrated in
\Fref{fig:Fig1}a. While graphene is planar atomic layer of carbon,
phosphorene is a non-planar layer of phosphorus atoms, forming a
puckered structure where atoms are located on two parallel planes. As
a result, phosphorene has an anisotropic crystal structure.

Figure \ref{fig:Fig1}b shows the projection of the phosphorene crystal
onto the a plane. The rectangular area indicates a unit cell, which
contains four atoms labeled $A$, $B$, $A'$ and $B'$. Their positions
in the unit cell are: $\bm{\tau_A}$=(u$c_0$,0,v$b_0$),
$\bm{\tau_B}=((1/2-u)c_0,a_0/2,vb_0)$,
$\bm{\tau_{A\textsc{\char13}}}=-\bm{\tau_{A}}$ and
$\bm{\tau_{B\textsc{\char13}}}=-\bm{\tau_{B}}$, where $a_0$ =
3.314\AA, $c_0$ = 4.376\AA, and $b_0$ = 10.48\AA, are the
corresponding lattice constants in $y$ (zigzag), $x$ (armchair) and
$z$ directions.\cite{YukHA1981,Osada15} Here, $u$ = 0.08056 and $v$ =
0.10168 are dimensionless crystal structure parameters. From these
atom locations, we can define the first eight lattice displacement
vectors in Table \ref{Tab:Tab1}.

\begin{table} [ht!]
\begin{ruledtabular}
\begin{tabular}{c c}
Order& Distances (\AA)\\\hline
  $\bm{d}_{1}=\bm{\tau}_{B}-\bm{\tau}_{A}$ &2.224  \\
  $\bm{d}^+_{2}=\bm{\tau}_{B'}+\bm{a}+\bm{c}-\bm{\tau}_{B}$, $\bm{d}^-_{2}=\bm{\tau}_{A'}-\bm{\tau}_{A}$ &2.244  \\
  $\bm{d}_{3}=\bm{a}$ &3.314  \\
  $\bm{d}_{4}=\bm{\tau}_{A}+\bm{a}+\bm{c}-\bm{\tau}_{B}$ &3.334  \\
  $\bm{d}_{5}=\bm{\tau}_{B'}+\bm{a}+\bm{c}-\bm{\tau}_{A}$ &3.475  \\
  $\bm{d}_{6}^{R}=\bm{\tau}_{B'}+\bm{2a}+\bm{c}-\bm{\tau}_{B}$, $\bm{d}_{6}^{L}=\bm{d}_{6}^{R}-4u\bm{c}$ &4.002  \\
  $\bm{d}^+_{7}=\bm{d}^-_{2}+\bm{c}$,  $\bm{d}^-_{7}=\bm{d}^+_{2}-\bm{c}$ &4.245 \\
  $\bm{d}_{8}=\bm{c}$ &4.376
\end{tabular}
\end{ruledtabular}
\caption{Intersite distances. Following Ref. \onlinecite{YukHA1981},
  the lattice vectors are defined as $\bm{a}$=$(0,a_0,0)$ and
  $\bm{c}$=$(c_0,0,0)$.}
\label{Tab:Tab1}
\end{table}

We include the 3$s$ and 3$p_{x,y,z}$ electrons in the partially filled
atomic shells and neglect any spin-orbit coupling since phosphorus is
a low-$Z$ element.\cite{LipAI14} The effective Hamiltonian is
represented as the following 16$\times$16 matrix within the basis
($A_{s}$, $A_{p_x}$, $A_{p_y}$, $A_{p_z}$, $B_{s}$, $B_{p_x}$,
$B_{p_y}$, $B_{p_z}$, $A'_{s}$, $A'_{p_x}$, $A'_{p_y}$, $A'_{p_z}$,
$B'_{s}$, $B'_{p_x}$, $B'_{p_y}$, $B'_{p_z}$):
\begin{widetext}
\begin{equation}
 H_{\rm mono}(\bf{k}) = \left[
\begin{array}{cccc}
\bm{T}_0+\bm{T}_3+\bm{T}_8 & \bm{T}_1+^t\bm{T}^*_4 & \bm{T}^{-}_2+\bm{T}^{L}_6+\bm{T}^{+}_7 &\bm{T}_5\\
^t\bm{T}_1^*+\bm{T}_4 & \bm{T}_0+\bm{T}_3 +\bm{T}_8 & \bm{T}_5 &\bm{T}^{+}_2+\bm{T}^{R}_6+\bm{T}^{-}_7\\
^t\bm{T}^{-*}_2+^t\bm{T}^{L*}_6+^t\bm{T}^{+*}_7 & ^t\bm{T}^*_5 & \bm{T}_0+\bm{T}_3 +\bm{T}_8 &^t\bm{T}_1^*+\bm{T}_4 \\
^t\bm{T}^*_5 & ^t\bm{T}^{+*}_2+^t\bm{T}^{R*}_6+^t\bm{T}^{-*}_7 & \bm{T}_1+^t\bm{T}^*_4 & \bm{T}_0+\bm{T}_3+\bm{T}_8
\end{array}
\right].
\label{Hamiltonianmoment}
\end{equation}
\end{widetext}

We take into account up to eighth nearest neighbor couplings (see
\Fref{fig:Fig1}b) through eight 4$\times$4 matrices referred to as
$\bm{T}_i$, within the $\{| m \rangle\}$ basis of atomic shells. Here,
the index $m$ represents $s$, $p_x$, $p_y$, and $p_z$ orbitals. The
interatomic matrix elements $T^{m,m'}_i(\bm{k})$ are given by the
expression
\begin{equation}
\label{Hamiltonian}
T^{m,m'}_i(\bm{k}) = {t}^i_{mm'}\sum\limits_{j=1}^N e^{
  i(\bm{R'_j+r_l'-r_l})\cdot \bm{k}},
\end{equation}
where $N$ is the number of unit cells, $R_j$ denotes the position of
the $j$th unit cell of the Bravais lattice, and $r_l$ is the position
of the atom $l$ within the unit cell. In this case, we sum only over
the adjacent unit cells $j$ which contain the atoms $l$, with the
displacement vector magnitude given by $|\bm{R'_j+r_l'-r_l}| =
|\bm{d}_i|$. The lattice displacement vector are provided in Table
\ref{Tab:Tab1}. The hopping amplitudes $\bm{t}^i_{mm'}$ are initially
written in terms of Slater-Koster (SK) parameters.\cite{SlaK54}

\subsection{Reference Density Function Band Structure}

In order to optimize the tight-binding model, we employ a DFT
calculation to generate a reference band structure for phosphorene. We
use the supercell method with a plane-wave basis set at a cutoff
energy of 500 eV and the projector-augmented wave technique,
\cite{bloechl994, Kresse1999} as implemented in the Vienna {\it
  ab-initio} Simulation Package (VASP).\cite{kresse1996b, kresse1993}
We use the hybrid HSE06\cite{Heyd2003, HSE2006} functional for the
exchange-correlation of the electrons. The supercell consists of a
unit cell of monolayer black phosphorus with experimental lattice
parameters, bond lengths, and bond angles \cite{Brown1965} and a
vacuum of 15 \AA. The Brillouin zone is sampled over a $(9 \times 12
\times 1)$ $k$-point mesh for a self-consistent calculation. The
electronic band structure along high-symmetry directions is calculated
with a finer mesh of $k$-points and then projected onto every orbital
of each atom to resolve the symmetry character of the corresponding
wavefunctions (i.e., their $l$ and $m$ numbers). The band structure
obtained in our DFT-HSE06 calculations shows that single layer black
phosphorus is a direct band gap material with a band gap ($E_g$) of
1.1 eV, which is quite close to the experimentally measured values so
far (of 1.0 and 1.55 eV).\cite{JinXZ14,SapWM14,LiuNZ14,WanJS15}

\subsection{Optimization of hopping parameters}

Our tight-binding model Hamiltonian has $16\times 16$ hopping
amplitudes $t^i_{mm'}$. Due to symmetry, we only need to calculate 58
of these elements. These parameters are optimized to reproduce the
main characteristics of the energy bands near the main gap, as
obtained from DFT-HSE06 calculations. The route to approximate the
band structure is the following:

\begin{itemize}

 \item Step 1: Following Slater and Koster,\cite{SlaK54} we initially
   constructed the tight-binding Hamiltonian for phosphorous $3s$,
   $3p_x$ , $3p_y$ and $3p_z$ orbitals (see
   \Eqref{Hamiltonianmoment}). Under this scheme, the hopping
   amplitudes $t^i_{mm'}$ are defined at first as a function of
   Slater-Koster parameters ($V_{ss\sigma}$, $V_{sp\sigma}$,
   $V_{pp\sigma}$ and $V_{pp\pi}$), as described in detail in the
   Appendix \ref{SK}. By diagonalizing $H_{\rm mono}({\bf k})$ for
   this first choice of hopping parameters, we obtain the band
   structure of monolayer phosphorene, as shown in
   Fig. \ref{fig:Fig2}, together with the band structure for the
   DFT-HSE06 calculations. Unfortunately, it is clear from
   Fig. \ref{fig:Fig2} that this simple model fails to resolve finer
   details in the band structure, which are important for electronic
   transport calculations. Although the results obtained from the SK
   parameters are largely inaccurate when compared with DFT-HSE06
   calculations, they serve as a useful starting point to optimize the
   tight-binding parameters, using the method of least squares as
   described in the next two steps.

\begin{figure} [h!]
\begin{center}
\includegraphics[width=1\columnwidth]{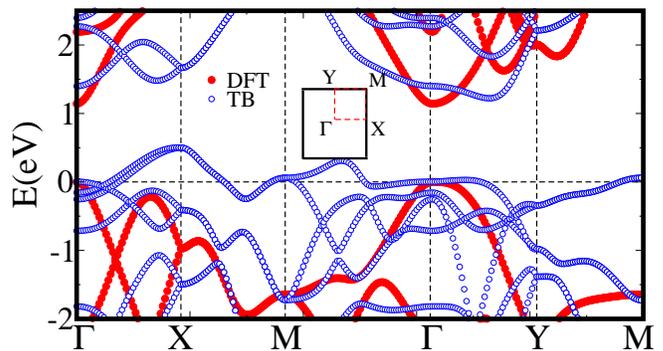}
\caption{(Color online) Comparison between the band structures
  obtained with the DFT-HSE06 (red squares) and with Slater-Koster
  tight-binding model (blue circles)}
\label{fig:Fig2}
\end{center}
\end{figure}

 \item Step 2: We then generate several different sets of parameters
   $t^i_{mm'}$ from the initial hopping amplitudes obtained in step 1.
   Each of these sets is generated by adding to the initial hoppings a
   random amplitude $\delta V$, taken from a uniform distribution over
   the interval $[-1,1]$meV. Following this, we take here 1000
   slightly different parameter sets.

 \item Step 3: For each of the new parameter sets, we choose the same
   number of representative $k$-points and calculate, by
   diagonalization of $H_{\rm mono}({\bf k})$, the corresponding band
   energies $E_n(k)$, where $n$ is the band index. We find the best
   tight-binding set of parameters among the 1000 generated by
   choosing the set that gives the lowest possible $\chi^2$ function,
   where $\chi^2$ is just a sum of weighted squared
   residuals,\cite{RidLRML15} namely,
\begin{equation}
\label{Hamiltonian}
\chi^2 = \sum \limits_{i=C,V} \sum \limits_{j=1}^N
\frac{\left[E_{i}^{TB}(j)-E_{i}^{DFT}(j)\right]^2}{\sigma_j^2},
\end{equation}
where $j$ labels the $k$ points and $i$ labels the lowest conduction
(C) and highest valence (V) energy bands. To improve the approximation
we give a larger weight $\sigma_j=1$ to points $\big(k,E_n(k)\big)$
near the $\Gamma$ point. In addition, we take a larger concentration
of points around $\Gamma$ to reproduce the effective band masses
around this high-symmetry point.

\item Step 4: Steps 2 and 3 are iteratively repeated (restarting step
  2 each time with the best set selected in step 3) until $\chi^2$
  becomes smaller than 1 meV$^2$. When this convergence criterion is
  satisfied, the optimized tight-binding parameters are obtained.
\end{itemize}

Table \ref{Tab:Tab2} presents the best fitting parameters we obtained
using the the optimization procedure described above.It is important 
to  emphasize that these parameters correspond to  the single layer black phosphorus, 
and, although they would be modified for other phosphorene allotropes,\cite{ZhuT2014}  
the same optimization procedure to find the best tight-binding parameters can be applied.


\begin{table*} [t!]
\begin{center}
\begin{ruledtabular}
\begin{tabular}{r r r r r r r r r r r}
       i   &$t^i_{ss}$  &$t^i_{sx}$     &$t^i_{sy}$     &$t^i_{sz}$     &$t^i_{xx}$     &$t^i_{xy}$     &$t^i_{xz}$     &$t^i_{yy}$     &$t^i_{yz}$     &$t^i_{zz}$     \\\hline
       1   &    1.402   &   -0.316      &    0.247      &                &    1.236      &    2.665      &               &    6.083      &               &   -1.770      \\

       2   &   -1.418   &   -1.173      &               &   -0.775      &   -1.541      &               &   -0.841      &   -5.809      &               &    2.170      \\

       3   &    0.349   &               &   -0.100      &               &    0.079      &               &               &    0.568      &               &    0.042      \\

       4   &   -0.239   &    0.300      &   -0.639      &               &    0.599      &    0.904      &               &    1.006      &               &    0.753      \\

       5   &   -0.255   &   -0.303      &   -0.246      &   -0.180      &    0.328      &   -0.038      &    0.166      &    0.654      &    0.659      &    0.096      \\

       6   &   -0.123   &    0.259      &   -0.072      &    0.100      &    0.063      &    0.305      &   -0.055      &   -0.206      &   -0.683      &   -0.313      \\

       7   &   -0.221   &   -0.146      &               &   -0.128      &    0.349      &               &   -0.077      &   -0.018      &               &    0.628      \\

       8   &    0.266   &   -0.260      &               &               &   -0.588      &               &               &    0.147      &               &   -0.037      \\
\end{tabular}
\end{ruledtabular}
\caption{Tight-binding model parameters obtained by optimization. The values are given in units of eV.}
\label{Tab:Tab2}
\end{center}
\end{table*}

\section{Band structure and orbital contribution}
\label{sec:bandstructure}

\begin{figure} [h!]
\begin{center}
\includegraphics[width=1\columnwidth]{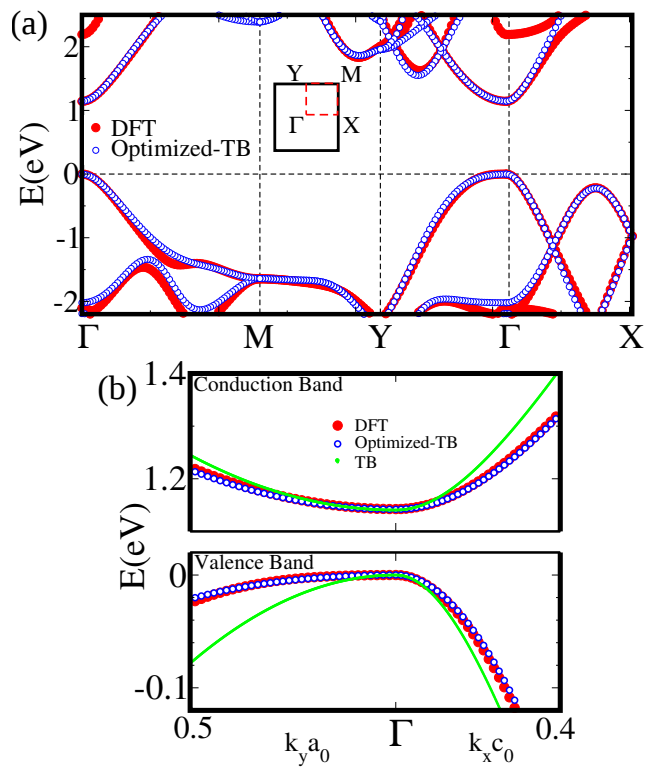}
\caption{(Color online) (a) Comparison between the band structures
  obtained with DFT-HSE06 (red squares) and with the optimized
  tight-binding model (blue circles). (b) Zoom around the $\Gamma$
  point, showing that the optimized model accurately reproduces the
  valence and conduction bands from DFT near the gap region. The green
  continuous line represents the tight-binding results considering
  only the $p_z$ orbital.}
\label{fig:Fig3}
\end{center}
\end{figure}

In Fig. \ref{fig:Fig3} we show a comparison between the band
structures for a single-layer phosphorene obtained with DFT-HSE06 and
that obtained from the optimized tight-binding model described in the
previous section. The optimized tight-binding model is in good
agreement with the DFT-HSE06 results and is quite accurate near the
minimum of the conductance band and the maximum of the valence band
(see Fig. \ref{fig:Fig3}a). These are the most important regions of
the spectrum as far as electronic transport is concerned and therefore
accuracy here is fundamental for obtaining realistic predictions for
transport properties.


We point out that this level of accuracy is missing in previous
studies, where simpler tight-binding Hamiltonians were employed; for
example, in models based on a single $p$ orbital.
\cite{RudYK2015,YuaRK2015,TagEZ2015,PopKS15,RodCC2014,RudK2014} In
those simpler models the bands near the main energy gap have a large
discrepancy with respect to the DFT results (the green continuous line
in \Fref{fig:Fig3}b represents the tight-binding results considering
only the $p_z$ orbital). For most of these previous studies, the focus
was in describing accurately only the main energy gap of the band
structure at the $\Gamma$ point. In contrast, our optimized
tight-binding model, in addition to capturing the energy gap, is able
to describe the bands structure in the $\Gamma \rightarrow Y$ and
$\Gamma \rightarrow X$ and $\Gamma \rightarrow M$ directions with high
accuracy, thus allowing us to properly study the effects of anisotropy
on transport properties.

Tight-binding methods employing orbitals $sp^3$
\cite{YukHA1981,Osada15} and $sp^3d^5$ \cite{LeeSQ16} have been
developed including up to second nearest neighbors. Those studies show
a clear deviation with respect to DFT results. A description of the
electronic structure of phosphorene supported by the Wannier functions
formalism has also been performed.\cite{JiwC15} This study was
successful in achieving a notable accuracy in the band structure of
phosphorene, but the computational cost would be too heavy for
studying electronic transport, where very large real-space lattices
are required.

\begin{figure*} [ht!]
\begin{center}
\includegraphics[width=1\columnwidth]{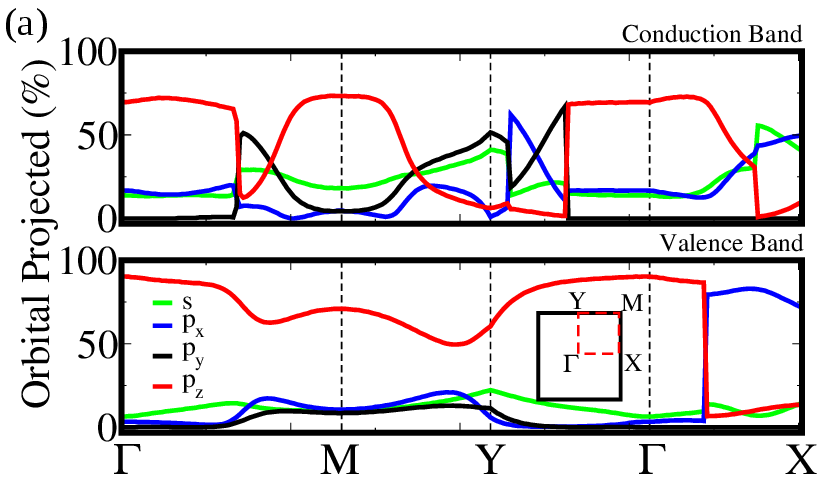}
\includegraphics[width=1\columnwidth]{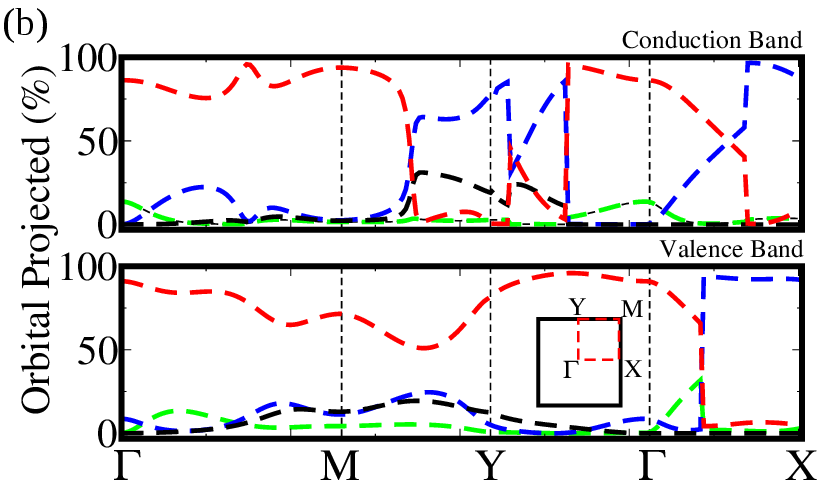}
\caption{ (Color online) (a) Orbital-projected band structure obtained
  with DFT-HSE06. (b) Orbital-projected band structure obtained with
  the optimized tight-binding model. The contribution of each orbital
  is shown by color: $s$(green), $p_x$(blue), $p_y$(black) and
  $p_z$(red).}
\label{fig:Fig4}
\end{center}
\end{figure*}

In \Fref{fig:Fig4} we show a comparison between the main orbital
composition obtained from DFT-HSE06 (\Fref{fig:Fig4}a) and from the
optimized tight-binding model (\Fref{fig:Fig4}b) near the valence band
maximum and the conduction band minimum. Around the $\Gamma$ point, it
can be seen that the main orbital contribution to both bands comes
from the $p_z$ orbitals (about 90\%). The orbital contributions around
the high-symmetry points $M$, $X$, $Y$, and $\Gamma$ from the
optimized tight-binding model show a qualitatively correct composition
of the orbitals for both conduction and valence bands when compared
with the DFT-HSE06 results. In particular, the composition of the
conduction band shows non-negligible contributions from $s$, $p_x$,
$p_y$ and $p_z$ orbitals.


\section{Anisotropy}
\label{sec:anisotropy}

\begin{figure} [h!]
\includegraphics[width=1\columnwidth]{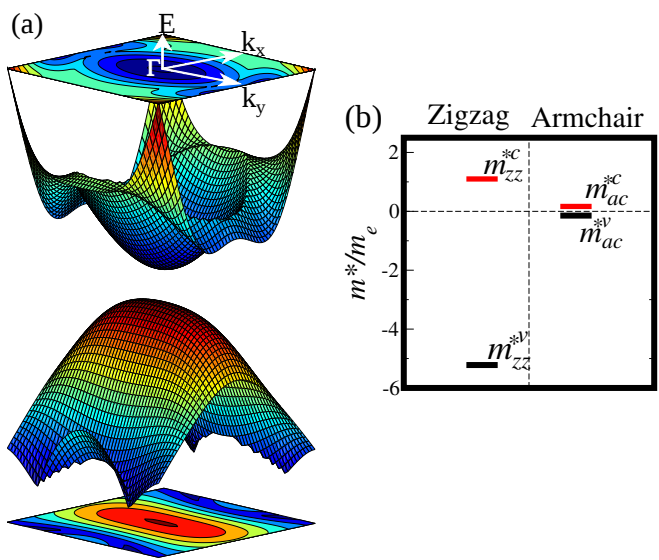}
\caption{ (a) Three-dimensional contour plot of the valence and
  conduction bands around the $\Gamma$ point. (b) Effective masses
  along armchair and zigzag directions. $m^{*c}$ (in red) and $m^{*v}$
  (in black) are the effective masses for the conduction and valence
  bands, respectively, making evident the anisotropy for both bands.}
\label{fig:Fig5}
\end{figure}

Figure \ref{fig:Fig5}a shows the dispersion of the valence and
conduction bands $E(\bm{k})$ around the $\Gamma$ point obtained
numerically by diagonalizing the tight-binding Hamiltonian in
Eq. (\ref{Hamiltonianmoment}). Both bands are clearly anisotropic, as
it can be seen in the top and bottom contours. One can see that the
valence band near the $\Gamma$ point is flatter along the $k_y$
direction than along the $k_x$ direction, implying that the hole
carriers moving along the zigzag direction are heavier than in the
armchair direction. A similar behavior is also observed for the
dispersion of conduction band. Strong anisotropy for both electron and
hole carriers was observed experimentally in multilayer
phosphorene.\cite{MisCY2015,SacMD16} The different effective masses of
the valence and conduction bands along the armchair and zigzag
directions is consistent with the in-plane anisotropy reported in
several transport experiments.\cite{XiaWJ14,RuL14,WanJS15} In this
paper we make this connection quantitative.

The anisotropy can be further identified directly from the anisotropic
effective masses as shown in \Fref{fig:Fig5}b. From our tight-binding
band structure we extract the effective masses for the electrons and
holes through the expression $m^* = \hbar^2 \left(\frac{\partial^2
  E}{\partial k^2}\right)^{-1}$. The resulting effective masses at
$\Gamma$ point along the armchair direction are $m^{*v}_{ac} = -0.1678
m_e$ and $m^{*c}_{ac} = 0.1990 m_e$ for holes and electrons
respectively. Here, $m_e$ is the free electron mass. The effective
masses along the zigzag direction are much heavier than armchair
direction: $m^{*v}_{zz} =-5.3525 m_e$ and $m^{*c}_{zz} = 0.7527 m_e$
for holes and electrons respectively. These values are also in close
agreement with other DFT calculations.\cite{JinXZ14} We note that the
single-orbital ($p_z$) tight-binding method (see green line in in
\Fref{fig:Fig2}c) cannot accurately capture this effective mass
anisotropy.


\section{Electronic properties approach: Anisotropic resistivity}
\label{sec:transport}

\subsection{Hamiltonian in real space}

Numerical studies of electronic transport in 2D materials have to
strike a compromise between the model complexity and the length scales
that can be investigated. Complex models requiring many basis states
per unit cell can only be used to investigate small systems, where the
diffusive regime common to experiments cannot be probed. Because of
their relative simplicity and small basis state sets, the use of
heuristic tight-binding models has grown in interest in the last
decade.\cite{YuaRK2015,LeeSQ16} Very large systems can be studied with
these models, sometimes involving over a billion
atoms,\cite{FerMu2016} in contrast to \textit{ab initio}
approaches. When the length scales associated to charge carrier
scattering involve more than a few lattice spacings, tight-binding
models are the only practical choice. We study transport properties of
phosphorene starting from our optimized tight-binding Hamiltonian in
$k$-space, \Eqref{Hamiltonianmoment}. The real-space tight-binding
Hamiltonian used in the numerical calculations, denoted by $H$
includes nearest-neighbor hopping terms (within the same unit cell),
as well as next-to-nearest-neighbor ones (between adjacent cells), as
discussed in Sec. \ref{sec:model}. Using second quantization, the
real-space Hamiltonian can be written as
\begin{equation}
\label{eq:2LBBD}
\begin{array}{l}
H = \displaystyle \sum_{i=1}^{\cal N} \sum_{\alpha}\left(
\varepsilon^{i}_{\alpha}\, c^{i\dagger}_{\alpha}\, c^i_{\alpha} +
\sum_{j} \sum_{\beta}t^{j}_{\alpha,\beta} c^{i\dagger}_{\alpha}\,
c^j_{\beta} \right) +\text{H.c.},
\end{array}
\end{equation}
where $i$ runs over the ${\cal N}$ lattice sites, $j$ runs over the
eight neighboring sites of $i$ and $\alpha$ and $\beta$ run over $s$
and $p$ orbitals. Here, $\varepsilon^{i}_{s} = -17.10$ eV and
$\varepsilon^{i}_{p} = -8.33$ eV are the energy levels of 3$s$ and
3$p$ orbitals of phosphorus, respectively.\cite{AbrGD63}
$t^j_{\alpha,\beta}$ is the hopping integral between the $i$th and its
$j$th neighbor, and $c^i_{\alpha}$ ($c^{i\dagger}_{\alpha}$) is the
annihilation (creation) operator of electrons at orbital $\alpha$ on
the site $i$. The different hopping and on-site terms can be
visualized in \Fref{fig:Fig1}b.

Although we use a relatively simple Hamiltonian to describe
phosphorene, it not only captures the physics qualitatively well, but
is also quantitatively approximately correct. This is because, in the
absence of disorder, both the energy bands and the wavefunctions near
the main gap closely resemble those calculated from an accurate
\textit{ab initio} theory. Nevertheless, we emphasize that the choice
of tight-binding parameters is not unique and not yet fully settled,
with several different parameter sets proposed in the
literature.\cite{YuaRK2015,LeeSQ16}

\subsection{Transport calculations}

\begin{figure} [ht!]
\begin{center}
\includegraphics[width=0.9\columnwidth]{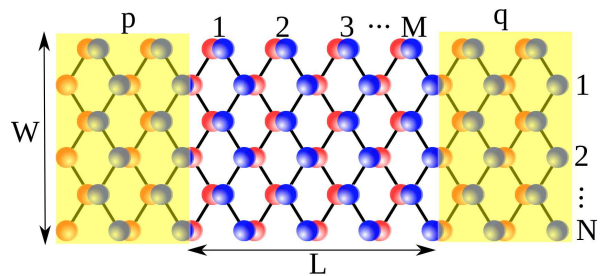}
\caption{(Color online) Schematic representation of a phosphorene
  sample of length $L$ and width $W$, and the corresponding number
  ($M$ and $N$) of unit cells in the armchair and zigzag
  directions. The shadow areas represent the left (p) and right (q)
  semi-infinite contacts.}
\label{fig:Fig6}
\end{center}
\end{figure}

Our calculations of the two-terminal linear conductance follow the
well-established Caroli formula,\cite{CarCLN71}
\begin{equation}
 {\cal T}(E) = \text{Tr}\left[ \Gamma_\text{p}\, G^r\,
   \Gamma_\text{q}\, G^a \right],
\end{equation}
which relates the transmission probability (transmittance) ${\cal
  T}(E)$ at a fixed carrier energy $E$ to the Green's functions $G^r$
and $G^a=(G^r)^{\dagger}$ of the sample when coupled to source (p) and
drain (q) contacts (represented by shadow areas in in
\Fref{fig:Fig6}).  The trace indicates a sum over all transverse
channels (or, equivalently, over all atomic sites at the
sample-electrode contact region). The matrices $\Gamma_\text{p,q}$
represent the imaginary part of the self-energy due to the coupling to
the electrodes, $\Gamma_{{p}({q})} = i\big[ \Sigma_{{p}({q})} -
  \Sigma_{{p}({q})}^\dag \big]$. The Green's functions are obtained by
a recursive technique where the sample is split into atomic transverse
slices.\cite{LewMuc13} We assume that the electrodes are identical
semi-infinite phosphorene strips with no disorder; the strip Green's
function, which is a fundamental ingredient in the recursive
technique, is obtained numerically using a standard decimation
method.\cite{LopLS85}.

A schematic representation of the system contact-sample-contact is
shown in \Fref{fig:Fig6}, where $L$ and $W$ are the length and the
width of the phosphorene sample considered. $M$ and $N$ indicated in
the figure are the number of unit cells in armchair and zigzag
directions, respectively. Therefore, for transport along the armchair
direction, as is the case represented in \Fref{fig:Fig6}, $L=Mc_0$ and
$W=Na_0$. If the transport is calculate along the zigzag direction,
then $L=Na_0$ and $W=Mc_0$.

Within the Landauer-B\"uttiker formalism, the linear conductance
${\cal G}(E)$ at a given energy $E$ is directly related to the
transmission function ${\cal T}(E)$ between the contacts as
\begin{equation}
  {\cal G}(E) = {\cal G}_0\, {\cal T}(E),
\end{equation}
where ${\cal G}_0 = 2e^2/h$. The linear resistance follows
straightforwardly from $R = 1/{\cal G} = R_0/{\cal T}$, where
$R_0=1/{\cal G}_0 = 12.5$ K$\Omega$. The resistivity is obtained as
usual, namely, $\rho = R\, W/L$.

In experiments, it is the carrier density $n$ rather than the carrier
energy $E$ that can be controlled by a back gate. Thus, in order to
explore how the resistivity $\rho$ behaves as a function of disorder
strength, we perform the calculations at fixed values of $n$. For a
given realization of disorder, the latter is obtained through the
relation
\begin{equation}
\label{eq:cchargedensity}
n(E) = \frac{1}{A} \int^{E}_{E_0} dE^\prime\, {\nu(E^\prime)},
\end{equation}
where $A = W\, L$ is the sheet area and $E_0$ is a reference energy
(either the top of the valence band of the bottom of the conductance
band). Note that in the conduction band, $E>E_0$ and therefore the
integral is over positive energies (electrons), while in the valence
band, $E<E_0$ and the integral is over negative energies
(holes). $\nu$ is the global density of states, which can be readily
obtained from the energy dependence of the scattering matrix $S$,
\begin{equation}
\nu(E) = -\frac{i}{2\pi} \mbox{Tr} \left( S^\dagger \frac{\partial
  S}{\partial E} \right).
\end{equation}
The scattering matrix $S$ is evaluated in terms of the retarded
Green's functions $G^r$,\cite{LeeF81,LewMuc13}
\begin{equation}
S_{ab}(E) = - \delta_{ab} + i \hbar \frac{\sqrt{v_a v_b}}{a_0}
\sum_{i}\sum_{j} \chi^*_a(i)\, G^r(i,j)\, \chi_b(j),
\end{equation}
where i and j run over the sites at the contacts p and q, where the
propagating channels $a$ and $b$ are defined, respectively. Here $a_0$
is the lattice constant and $v_{a,b}$ and $\chi_{a,b}$ are,
respectively, the longitudinal propagation velocity and the transverse
wavefunction in the propagating channel. We obtain $v_{a,b}$ and
$\chi_{a,b}$ from the eigenvalues and eigenfunctions of the
$\Gamma_{p(q)}$ matrices,

\begin{equation}
\Gamma_{p(q)}(i,j)=\sum_{a}\chi_{a}(i)\frac{\hbar v_a}{a_0}\chi_{a}^*(j).
\end{equation}

\subsection{Disorder effects over the anisotropy}

We studied the disorder effects by using the optimized tight-binding
method previously described, which allows for very efficient
large-scale calculations of linear transport properties. To model
disorder, a superposition of Gaussian potential fluctuations is added
to the Hamiltonian in Eq. (\ref{eq:2LBBD}) as a diagonal
term,\cite{RycTB07}
\begin{equation}
\label{eq:cchargedensity}
U({\bf r}_i) = \sum \limits_{k=1}^{N_{\rm imp}} U_k\, e^{-|{\bf
    r}_i-{\bf R}_k|^2/2\xi^2},
\end{equation}
where $r_i$ denotes a lattice site. The $N_{\rm imp}$ Gaussian
scatterers have a width $\xi$, are located at random sites $\{{\bf
  R}_k\}_{k=1,N_{\rm imp}}$ drawn uniformly and have amplitudes
$\{U_k\}_{k=1,N_{\rm imp}}$ taken from a flat distribution in the
interval $\left[-\delta U/2, \delta U/2 \right]$. Let $n_{\rm imp} =
N_{\rm imp}/{\cal N}$ denote the density of scatterers. Motivated by
the two prevailing scattering mechanisms in phosphorene transistors,
we consider two extreme cases: (i) dense disorder ($n_{\rm imp}=1\%$)
with low amplitude of the Gaussian potential fluctuations
($0.03\leqslant \delta U \leqslant 0.14$ eV); and (ii) dilute disorder
($n_{\rm imp}=0.1\%$), with higher amplitudes of the Gaussian
potential ($0.2\leqslant \delta U \leqslant 2$ eV). Case (i) models
contaminants such as water, which attach to phosphorene by weak van
der Waals interactions (therefore the low amplitudes). Case (ii)
models background potential inhomogeneities like those caused by
screened charges in the substrate. Notice that although the Gaussian
potential we consider is short-range on the system-size scale
(correlation-length $\xi=1.5a_0$), it varies smoothly on the atomic
scale, corresponding to an effective disorder which mimics the effect
of screened charges from the substrate.\cite{RycTB07,LewMuc13}

\begin{figure*} [ht!]
\begin{center}
\includegraphics[width=2\columnwidth]{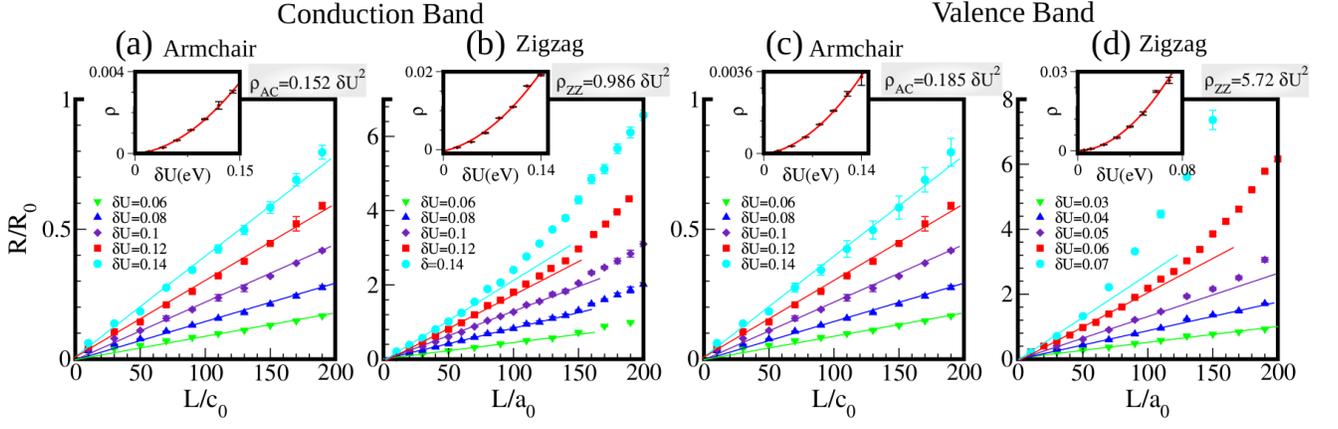}
\caption{(Color online) Average resistance as a function of length $L$
  for disorder case (i), corresponding to a dense concentration of
  scatterers ($n_{\rm imp}=1\%$), with low disorder amplitudes $\delta
  U$.  Continuous lines are linear fittings used to extract the
  resistivity in the diffusive regime of the data for each
  curve. Insets: Resistivity as a function of disorder amplitudes
  $\delta U$, showing a quadratic dependence.}
\label{fig:Fig7}
\end{center}
\end{figure*}

\begin{figure*} [ht!]
\begin{center}
\includegraphics[width=2\columnwidth]{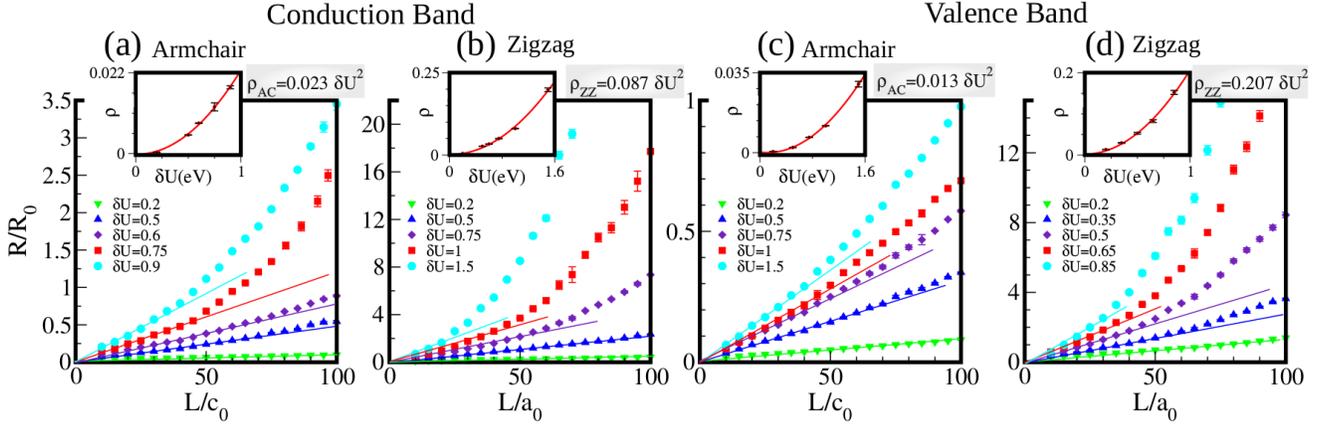}
\caption{(Color online) (Color online) Average resistance as a
  function of length $L$ for disorder case (ii), corresponding to a
  dilute concentration of scatterers ($n_{\rm imp}=0.1\%$), with high
  disorder amplitudes $\delta U$. Continuous lines are linear fittings
  used to extract the resistivity in the diffusive regime of the data
  for each curve. Insets: Resistivity as a function of disorder
  amplitudes $\delta U$, showing a quadratic dependence.}
\label{fig:Fig8}
\end{center}
\end{figure*}

In \Fref{fig:Fig7} we show the average resistance as a function of
length of the system for the disorder case (i). Different panels
correspond to the resistance along armchair or zigzag directions, for
conduction and valence bands, as indicated. In each one, we show
curves for different disorder potential amplitude $\delta U$. The
range of the Gaussian potential considered is $\xi=1.5a_0$ for all of
them. The average is computed over 500 disorder configurations and for
a carrier density $n = 3 \times 10^{12}$ cm$^{2}$, which brings the
Fermi energy close to the bottom (top) of the conduction (valence)
band. We present similar data for the disorder case (ii) in
Fig. \ref{fig:Fig8}

At zero temperature, the resistance strongly fluctuates from one
realization to another, which is typical for a coherent
quasi-one-dimensional system.
However, it is clear from Figs. \ref{fig:Fig7} and \ref{fig:Fig8} that
the linear behavior, which is characteristic from a diffusive regime,
is kept for longer lengths for the armchair direction than in the
zigzag direction, particularly when the amplitude $\delta U$ of the
disorder potential is increased.
For longer lengths, the average resistance increases much more rapidly
with length, marking the onset of strong localization. For both the
conduction and valence regions, we have found that a strong
localization regime sets in with increasing $L$, with an exponential
increase of the resistance.

In the diffusive regime, we can extract the resistivity $\rho$ for
different disorder amplitudes $\delta U$ from the linear fittings
indicated in Figs. \ref{fig:Fig7} and \ref{fig:Fig8}. The inset in
each graph shows the resistivity as function of $\delta U$, where we
find a very good match to a quadratic dependence for all cases. The
resulting quadratic fitting for the resistivity as a function of
$\delta U$ is indicated in the top right of each graph in
Figs. \ref{fig:Fig7} and \ref{fig:Fig8}. $\rho_{ZZ}$ and $\rho_{AC}$
indicates the resistivity along the zigzag and armchair directions,
respectively.

Using classical kinetic transport theory, the resistivity $\rho$ can
be related to the effective masses $m^*$ and the mean scattering time
$\tau$ through: $\rho=m^*/\tau n q^2$, where $n$ and $q$ are the
density and the charge of the carriers, respectively. Even though an
expression for $\tau$ is not exactly know, it can be estimated in
perturbation theory to be inversely proportional to $\delta
U^2$. Thus, the quadratic dependence on $\delta U$ we observe in the
data can be attributed to $\tau$.

It is reasonable to expect the resistivity to be anisotropic,
considering the anisotropy in the effective masses. If the classical
kinetic transport theory is applicable here, we would expect
$\rho_{ZZ}/\rho_{AC}=m^*_{ZZ}/m^*_{AC}$. In Table \ref{Tab:Tab3} we
summarize our results for the ratios of resistivities
$\rho_{ZZ}/\rho_{AC}$ for the two disorder cases analyzed and also for
two sample widths: a thinner one, with $60$ unit cells in width and a
larger one, with $150$ (where all other parameters are kept constant).
These results should be compared with the ratios of effective masses
obtained from the ordered phosphorene system band structure (see
Sec.\ref{sec:anisotropy}): $m^\ast_{ZZ}/m^\ast_{AC} = 6.6$ for the
conduction and $m^\ast_{ZZ}/m^\ast_{AC}=39.4$ for the valence band.
Our intention is to observe how different densities and amplitudes of
disorder change anisotropy.  First of all, we observe from the results
in Table \ref{Tab:Tab3} that increasing the width of the phosphorene
sample considered from 60 to 150 unit cells does not change
considerably the resistivity ratios, which means that we do not have
system size effects masking our results here.

Comparing the two disorder cases considered here, we can conclude from
the resistivity ratios in Table \ref{Tab:Tab3} that the higher
amplitudes of the disorder in the second case (even considering the 10
times lower concentration of scatterers) cause stronger impact in
diminishing the anisotropy when compared to the first
case. Nevertheless, in both cases the anisotropy is still evident and
in ratios that would be experimentally detected.

It is helpful to analyze the results in light of the product $\delta
U^2$ $\times$ $n_{imp}$, considering that in the Boltzmann transport,
mobility depends on this product.\cite{RycTB07,LewMuc13} In our
calculations, the impurity density $n_{imp}$ is kept fixed for each
disorder case, while $\delta U$ is varied. For the case with dense
disorder and low amplitude (shown in \Fref{fig:Fig7}), the value of
the product was mostly higher than for the case with dilute disorder
and high amplitude (show in \Fref{fig:Fig8}): $\delta U^2$ $\times$
$n_{imp}$ varies in the interval $[9\times10^{-7},2\times10^{-4}]eV^2$
for the former and in the interval
$[4\times10^{-5},4\times10^{-3}]eV^2$ for the latter. This is
consistent with the results summarized in Table \ref{Tab:Tab3}, where
one can observe the stronger suppression of the anisotropy for the
disorder with the higher value of the product $\delta U^2$ $\times$
$n_{imp}$. Interestingly, cases with different disorder type but with
the same $\delta U^2$ $\times$ $n_{imp}$ product values, present the
same resistance values in Figs. \ref{fig:Fig7} and \ref{fig:Fig8},
confirming the universality related to this product.


\begin{table} [ht!]
\begin{center}
\begin{ruledtabular}
\begin{tabular}{c  c c c}
               &   \footnotesize{Ordered}        &  \footnotesize{Dense disorder}           &\footnotesize{Dilute  disorder}  \\
               &  \footnotesize{ System}                        &\footnotesize{  Low amplitude }           & \footnotesize{High  amplitude}   \\
               & $m^\ast_{ZZ}/m^\ast_{AC}$ &  $\rho_{ZZ}/\rho_{AC}$    & $\rho_{ZZ}/\rho_{AC}$   \\ \hline
\footnotesize{Thinner sample}        &                           &                           &      \\
\rowcolor{gray!25}[0.95\tabcolsep]\footnotesize{Conduction }    &       6.6                 &    $6.5\pm 0.1$           &   $3.8\pm 0.1$ \\
\footnotesize{Valence }       &  39.4                     &      $31\pm 1$            &   $16\pm 1$\\\hline

\footnotesize{Larger sample}   &           &                &     \\
\rowcolor{gray!25}[0.95\tabcolsep]\footnotesize{Conduction}   &    6.6     &     $6.6\pm 0.1$   &  $4.1\pm 0.1$  \\
\footnotesize{Valence}      &    39.4        &     $32\pm 1$  &  $18\pm 1$  \\

\end{tabular}
\end{ruledtabular}
\caption{Ratio between the resistivity along the zigzag and armchair
  directions $\rho_{zz}/\rho_{ac}$ for dilute and dense disorder
  cases. Effective mass ratios are $m^\ast_{ZZ}/m^\ast_{AC}=6.6$ and
  $m^\ast_{ZZ}/m^\ast_{AC}=39.4$ for conduction and valence bands,
  respectively.}
\label{Tab:Tab3}
\end{center}
\end{table}

\section{Summary and conclusion}
\label{conclusion}

We have developed a tight-binding model for monolayer phosphorene that
accurately describes both conduction and valence band dispersions near
the Gamma point and approximates well the band compositions. The
additional accuracy came at the expense of introducing $s$ in addition
to $p$ orbitals, as well as hopping amplitudes involving eight
neighbors in total.

We optimized the model parameters by using as benchmark the electronic
structure obtained by density functional theory calculation based on
the HSE06 exchange-correlation functional. An excellent match between
effective masses near the main band gap and along major symmetry
directions was obtained.

Using the optimized tight-binding model and a recursive Green's
function technique, we computed the resistivity in the presence of
disorder for two relevant situations, which mimic two commons types of
disorder in phosphorene: (i) weakly bonded adsorbates (simulated by a
dense concentration of scatterers, with low amplitudes of the Gaussian
potential fluctuations), and (ii) screened charge traps in the
substrate (simulated by a dilute concentration of Gaussian correlated
disorder, with higher amplitudes), We found that the band mass
anisotropy is strongly manifest in the resistivity for the first
disorder case, where the ratio of the resistivity along zigzag and
armchair directions matches quite closely the ratio for the
corresponding effective masses. The anisotropy is weaker, but still
robust, in the second disorder case. Thus, we conclude that the most
prevailing types of disorder likely to be found in monolayer
phosphorene should not wash away the intrinsic band structure
anisotropy of this material. Transport experiments performed with
thick films of black phosphorus (which is a multilayer phosphorene),
have already demonstrated intrinsic
anisotropy.\cite{LiuNZ14,LiuLR16,NarAT83,YuiSS83} Based on our
results, we expect a similar behavior for monolayer systems.

\section*{Acknowledgments}
CJP and ALCP acknowledge financial support from FAPESP through Grant
2015/12974-5. Part of the numerical simulations were performed at the
computational facilities at CENAPAD-SP, UNICAMP and UCF Advanced
Research Computing Center. DFT calculations are supported in part by
the DOE grant DE-FG02-07ER46354.
\appendix

\section{Slater-Koster}
\label{SK}

We develop an effective tight-binding model based on the LCAO
method\cite{Osada15} and use DFT calculations as the basis for
adjusting the model parameters. We begin with a simplified LCAO model.
The hopping amplitudes depend of the transfer integral between two
adjacent atoms. The transfer integrals are given by $V_{ll'm}(d) =
\eta_{ll'}\hbar^2 /m_ed^2$, where $d$ is the inter-atomic distance,
$m_e$ is the electron rest mass, $l$ and $l'$ are the orbital
azimuthal quantum numbers $(s, p)$ of two atoms and $m$ is the common
orbital magnetic quantum number ($\sigma$, $\pi$).  $\eta_{ll'm}$ is a
dimensionless quantity that depends on the crystal structure. For the
simplified model, the parameters employed are: $\eta_{ss\sigma} =
-1.40$, $\eta_{sp\sigma} = 1.84$, $\eta_{pp\sigma} = 3.24$, and
$\eta_{pp\pi} = -0.81$.\cite{AbrGD63}

When expressed in momentum space, the tight-binding Hamiltonian is a
$16 \times 16$ matrix, (see \Eqref{Hamiltonianmoment}). Here the
elements $T_i$ represent $4\times 4$ matrices.  The $T_0$ matrix on
the diagonal expresses the energies of the four atomic sites:
\begin{equation}
 \bm{T}_0 = \left[
\begin{array}{cccc}
 \varepsilon_s&  0  & 0   & 0\\
  0  &\varepsilon_p &  0  &  0\\
  0  &  0  & \varepsilon_p& 0\\
  0  &  0  &  0  &\varepsilon_p
\end{array}
\right]
\end{equation}
Here, $\varepsilon_s$ = -17.10 eV and $\varepsilon_p$ = -8.33 eV
represent the energy levels of the 3$s$ and 3$p$ orbitals of
phosphorus, respectively. The nearest and next-nearest neighbor
coupling between atoms are represented by $T_1$ to $T_8$ respectively:
%
\begin{equation}
 \bm{T}_1 = \left[
\begin{array}{cccc}
 t^{(1)}_{ss}g^+_1& t^{(1)}_{sx}g^+_1  & t^{(1)}_{sy}g^-_1   & 0\\
 -t^{(1)}_{sx}g^+_1  &t^{(1)}_{xx}g^+_1 &  t^{(1)}_{xy}g^-_1  &  0\\
 -t^{(1)}_{sy}g^-_1  & t^{(1)}_{xy}g^-_1  & t^{(1)}_{yy}g^+_1& 0\\
  0  &  0  &  0  &t^{(1)}_{zz}g^+_1
\end{array}
\right],
\end{equation}
with
\begin{equation}
g^{\pm}_1(\bm{k})=e^{i\bm{d}_1\cdot \bm{k}}(1\pm e^{-\bm{a}\cdot \bm{k}});
\end{equation}
%
\begin{equation}
 \bm{T}^{\pm}_2 = \left[
\begin{array}{cccc}
 t^{(2)}_{ss}g^{\pm}_2&\pm t^{(2)}_{sx}g^{\pm}_2  &  0  & t^{(2)}_{sz}g^{\pm}_2\\
 \mp t^{(2)}_{sx}g^{\pm}_2  &t^{(2)}_{xx}g^{\pm}_2 &  0 & \pm t^{(2)}_{xz}g^{\pm}_2\\
  0&0_2  & t^{(2)}_{yy}g_2^{\pm}& 0\\
   -t^{(2)}_{sz}g^{\pm}_2  & \pm t^{(2)}_{xz}g^{\pm}_2 &  0  &t^{(2)}_{zz}g^{\pm}_2
\end{array}
\right],
\end{equation}
with
\begin{equation}
g^{\pm}_2(\bm{k})=e^{i\bm{d}^{\pm}_2\cdot \bm{k}};
\end{equation}
%
\begin{equation}
 \bm{T}_3 = \left[
\begin{array}{cccc}
 t^{(3)}_{ss}g^{+}_3      &  0  & t^{(3)}_{sy}g^{-}_3  & 0\\
 0  &t^{(3)}_{xx}g^{+}_3 &  0 & 0\\
 -t^{(3)}_{sy}g^{-}_3 & 0&t^{(3)}_{yy}g^{+}_3 & 0\\
   0  & 0 & 0  &t^{(3)}_{zz}g^{+}_3
\end{array}
\right],
\end{equation}
with
\begin{equation}
g^{\pm}_3(\bm{k})=e^{i\bm{d}_3\cdot \bm{k}}\pm e^{-i\bm{d}_3\cdot \bm{k}};
\end{equation}
%
\begin{equation}
 \bm{T}_4 = \left[
\begin{array}{cccc}
 t^{(4)}_{ss}g^+_4& t^{(4)}_{sx}g^+_4  & t^{(4)}_{sy}g^-_4   & 0\\
 -t^{(4)}_{sx}g^+_4  &t^{(4)}_{xx}g^+_4 &  t^{(4)}_{xy}g^-_4  &  0\\
 -t^{(4)}_{sy}g^-_4  & t^{(4)}_{xy}g^-_4  & t^{(4)}_{yy}g^+_4& 0\\
  0  &  0  &  0  &t^{(4)}_{zz}g^+_4
\end{array}
\right],
\end{equation}
with
\begin{equation}
g^{\pm}_4(\bm{k})=e^{i\bm{d}_4\cdot \bm{k}}(1\pm e^{-i\bm{a}\cdot \bm{k}});
\end{equation}
%
\begin{equation}
 \bm{T}_5 = \left[
\begin{array}{cccc}
  t^{(5)}_{ss}g^{+++}_5  & t^{(5)}_{sx}g^{+-+}_5  &  t^{(5)}_{sy}g^{-+-}_5  & t^{(5)}_{sz}g^{+++}_5\\
 -t^{(5)}_{sx}g^{+-+}_5  & t^{(5)}_{xx}g^{+++}_5  &  t^{(5)}_{xy}g^{---}_5  &  t^{(5)}_{xz}g^{+-+}_5\\
 -t^{(5)}_{sy}g^{-+-}_5  & t^{(5)}_{xy}g^{---}_5  &  t^{(5)}_{yy}g^{+++}_5  & t^{(5)}_{yz}g^{-+-}_5\\
 -t^{(5)}_{sz}g^{+++}_5  & t^{(5)}_{xz}g^{+-+}_5  &  t^{(5)}_{yz}g^{-+-}_5  &t^{(5)}_{zz}g^{+++}_5
\end{array}
\right],
\end{equation}
with
\begin{equation}
\begin{array}{c}
g^{+++}_5(\bm{k})=e^{i\bm{d}_5\cdot \bm{k}}[1+ e^{-i\bm{a}\cdot \bm{k}}+e^{-i\bm{c}\cdot \bm{k}}(1+e^{-i\bm{a}\cdot \bm{k}})]\\
g^{+-+}_5(\bm{k})=e^{i\bm{d}_5\cdot \bm{k}}[1+ e^{-i\bm{a}\cdot \bm{k}}-e^{-i\bm{c}\cdot \bm{k}}(1+e^{-i\bm{a}\cdot \bm{k}})]\\
g^{-+-}_5(\bm{k})=e^{i\bm{d}_5\cdot \bm{k}}[1- e^{-i\bm{a}\cdot \bm{k}}+e^{-i\bm{c}\cdot \bm{k}}(1-e^{-i\bm{a}\cdot \bm{k}})]\\
g^{---}_5(\bm{k})=e^{i\bm{d}_5\cdot \bm{k}}[1- e^{-i\bm{a}\cdot \bm{k}}-e^{-i\bm{c}\cdot \bm{k}}(1-e^{-i\bm{a}\cdot \bm{k}})]
\end{array};
\end{equation}
%
%
\begin{equation}
 \bm{T}^R_{6} = \left[
\begin{array}{cccc}
  t^{(6)}_{ss}g^+_{6R}   & t^{(6)}_{sx}g^+_{6R}  & t^{(6)}_{sy}g^-_{6R}   & t^{(6)}_{sz}g^+_{6R}\\
 -t^{(6)}_{sx}g^+_{6R}  &t^{(6)}_{xx}g^+_{6R}   &  t^{(6)}_{xy}g^-_{6R}  & t^{(6)}_{xz}g^+_{6R}\\
 -t^{(6)}_{sy}g^-_{6R}  & t^{(6)}_{xy}g^-_{6R}  & t^{(6)}_{yy}g^+_{6R}   & t^{(6)}_{yz}g^-_{6R}\\
 -t^{(6)}_{sz}g^+_{6R}   & t^{(6)}_{xz}g^+_{6R}  & t^{(6)}_{yz}g^+_{6R}  & t^{(6)}_{zz}g^+_{6R}
\end{array}
\right],
\end{equation}
with
\begin{equation}
g^{\pm}_{6R}(\bm{k})=e^{i\bm{d}_{6R}\cdot \bm{k}}(1\pm e^{-i2\bm{a}\cdot \bm{k}});
\end{equation}
\begin{equation}
 \bm{T}^L_{6} = \left[
\begin{array}{cccc}
  t^{(6)}_{ss}g^+_{6L}  & -t^{(6)}_{sx}g^+_{6L}  & t^{(6)}_{sy}g^-_{6L}   & t^{(6)}_{sz}g^+_{6L}\\
  t^{(6)}_{sx}g^+_{6L}  & t^{(6)}_{xx}g^+_{6L}  & -t^{(6)}_{xy}g^-_{6L}  &-t^{(6)}_{xz}g^+_{6L}\\
 -t^{(6)}_{sy}g^-_{6L}  & -t^{(6)}_{xy}g^-_{6L}  & t^{(6)}_{yy}g^+_{6L}   & t^{(6)}_{yz}g^-_{6L}\\
 -t^{(6)}_{sz}g^+_{6L}  &-t^{(6)}_{xz}g^+_{6L}  & t^{(6)}_{yz}g^+_{6L}  & t^{(6)}_{zz}g^+_{6L}
\end{array}
\right],
\end{equation}
with
\begin{equation}
g^{\pm}_{6L}(\bm{k})=e^{i\bm{d}_{6L}\cdot \bm{k}}(1\pm e^{-i2\bm{a}\cdot \bm{k}});
\end{equation}
%
%
\begin{equation}
 \bm{T}^{\pm}_7 = \left[
\begin{array}{cccc}
 t^{(7)}_{ss}g^{\pm}_7      &\pm t^{(7)}_{sx}g^{\pm}_7  &   0                & t^{(7)}_{sz}g^{\pm}_7\\
 \mp t^{(7)}_{sx}g^{-}_7      &    t^{(7)}_{xx}g^{\pm}_7  &   0                &\pm t^{(7)}_{xz}g^{\pm}_7\\
 0                          &   0                    &t^{(7)}_{yy}g^{\pm}_7 & 0\\
  -t^{(7)}_{sz}g^{\pm}_7      &\pm t^{(7)}_{xz}g^{\pm}_7   & 0                  &t^{(7)}_{zz}g^{\pm}_7
\end{array}
\right],
\end{equation}
witj
\begin{equation}
g^{\pm}_7(\bm{k})=e^{i\bm{d}^{\pm}_7\cdot \bm{k}};
\end{equation}
%
%
\begin{equation}
 \bm{T}_8 = \left[
\begin{array}{cccc}
 t^{(8)}_{ss}g^{+}_8      &   t^{(8)}_{sx}g^{-}_8  & 0  & 0\\
  -t^{(8)}_{sx}g^{-}_8  &t^{(8)}_{xx}g^{+}_8 &  0 & 0\\
 0 & 0&t^{(8)}_{yy}g^{+}_8 & 0\\
0 &0&0 &t^{(8)}_{zz}g^{+}_8
\end{array}
\right],
\end{equation}
with
\begin{equation}
g^{\pm}_8(\bm{k})=e^{i\bm{d}_8\cdot \bm{k}}\pm e^{-i\bm{d}_8\cdot \bm{k}}.
\end{equation}
In those relations, $\bm{t}^{i}_{ss} = V_{ss\sigma}({d}_{i})$,
$\bm{t}^{i}_{\alpha\beta} =
(d_{i}^{\alpha}d_{i}^{\beta}/(d_{i})^2)V_{pp\sigma}({d}_{i}) +
(\delta_{\alpha\beta}
-d_{i}^{\alpha}d_{i}^{\beta}/(d_{i})^2)V_{pp\pi}({d}_{i})$ and
$\bm{t}^{i}_{s\alpha} = (d_{i}^{\alpha}/d_{i})V_{sp\sigma}({d}_{i})$,
where $\bm{d}_{i} = (d_{i}^{x},d_{i}^{y},d_{i}^{z})$ and ${d}_{i} =
|\bm{d}_{i}|$. The indices run as follows: $i=1,\ldots,8$ and
$\alpha,\beta$=x,y,z. The phase factors $g_i$ are defined as function
of the distances and the wave number $\bf{k}$.

These definitions are similar to those used in the previous models in
the literaturel,\cite{Osada15} with the addition of new interatomic
matrix elements $T_3$, $T_4$, $T_5$, $T_6$, $T_7$, and $T_8$. The
reason for introducing these new parameters is that the two
interatomic matrix elements $T_1$ and $T_2$ provided by the
Slater-Koster coefficients are not sufficient to accurately describe
the band structure of phosphorne. These must be modified in order to
provide an accurate representation of the band gap. By diagonalizing
$H$, the band dispersion of monolayer phosphorus can be obtained, as
shown in \Fref{fig:Fig2}.

\bibliography{bibliograph}

\end{document}